\documentclass[multphys,vecphys]{svmult}

\usepackage{makeidx}         % allows index generation
\usepackage{graphicx,subfigure,epsf}        % standard LaTeX graphics tool
\usepackage{multicol}        % used for the two-column index
\usepackage[bottom]{footmisc}% places footnotes at page bottom
\def\xmm{{\it XMM-Newton\/}}
\def\cha{{\it Chandra\/}}

\def \h50 {$h_{50}$}
\def \h70 {$h_{70}$}

\begin{document}

\title*{Radio Source Heating in the ICM: The Example of Cygnus A}
\author{Elena Belsole\inst{1}\and
Andrew C. Fabian \inst{1,2}}
\institute{Institute of Astronomy, University of Cambridge, Madingley Road, Cambridge CB3 0HA, UK, 
\texttt{elena@ast.cam.ac.uk}
\and \texttt{acf@ast.cam.ac.uk}}
\maketitle
\begin{abstract} One of the most promising solutions for the cooling flow problem involves energy injection
from the central AGN.
However it is still not clear how collimated jets can heat the ICM at 
large scale, and very little is known concerning the effect of radio lobe expansion as 
they enter into pressure equilibrium with the surrounding cluster gas. Cygnus A is one 
of the best examples of a nearby powerful radio galaxy for which the synchrotron emitting
plasma and thermal emitting intracluster medium can be mapped in fine detail, and previous
observations have inferred possible shock structure at the location of the cocoon.
We use new \xmm\ observations of Cygnus A,
in combination with deep \cha\ observations, to measure the temperature of the intracluster medium 
around the expanding radio cavities. We investigate how inflation of the 
cavities may relate to shock heating of the intracluster gas, and whether such a mechanism 
is sufficient to provide enough energy to offset cooling to the extent observed. 
\end{abstract}

\section{Introduction}
\label{introbel}
% Always give a unique label
% and use \ref{<label>} for cross-references
% and \cite{<label>} for bibliographic references
% use \sectionmark{}
% to alter or adjust the section heading in the running head
Cygnus A (3C\,405) is the most radio-luminous Active Galactic Nucleus (AGN) to a redshift of $\sim 1$, and is the third-brightest source in the radio sky. Its power ($L_{\rm 178 MHz} = 6\times10^{27}$ W Hz sr$^{-1}$) and its proximity ($z=0.0562$) have made it one of the most studied extra-galactic sources.

In X-rays Cygnus A was first observed with the {\em Uhuru} satellite. Later {\em Einstein Observatory} \cite{arnaud84} and {\em ROSAT} observations \cite{hcp94,rf96}  have shown that the X-ray emission is dominated by the thermal radiation from  intra-cluster medium (ICM) and found that the gas in the inner 50 kpc is significantly cooler than the ICM at larger scale, with an inferred cooling flow rate of $\sim250$ M$_{\odot}$ yr$^{-1}$. 

The large scale temperature distribution of the ICM was mapped with 
{\em ASCA} \cite{msv99}, and suggests a merging event between two similar clusters, 
the one hosting Cyg A and a secondary cluster detected to the north-west (NW), at a projected distance of 11.5 arcmin ($\sim 740 h^{-1}$ kpc) .  
The merger scenario is in agreement with the galaxy distribution (\cite{owenetal97,lom05}).

More recently, Cygnus A has been studied with \cha.
The jet and hotspots are clearly detected and coincide with the radio.
The magnetic field properties of the hotspots of
the radio galaxy were studied by 
\cite{wys00}, while \cite{youngetal02} focused on the X-ray
emission from the nucleus. The large scale structure and the interaction between the 
radio source and the ICM was discussed by \cite{smithetal02} and \cite{wsy06}. The central 2.5 arcmin ($\sim 160 h^{-1}$ kpc) 
are characterised by a filamentary structure of spiral-like shape with the nucleus at its centre (named ``belts'' by \cite{smithetal02}).  \cha\ also detects, 
with a finer resolution than {\em ROSAT} HRI, the edge-brightened emission around the cavity coincident 
with the radio lobes (\cite{carillietal94}). This elliptically shaped structure encompassing the whole radio source has a major axis of $\sim 1.1$ 
arcmin ($\sim 70 h^{-1}$ kpc) coincident with the direction of the radio jet, and has been interpreted as the cocoon of shocked gas due to the expansion of the radio galaxy (\cite{wsy06}).

In this paper we present new \xmm\ data of Cygnus A. The larger field
of view (FoV) and sensitivity of \xmm\ allow us to reveal new features
in the X-ray emission from the cluster environment of this object. We
discuss  the effect of the radio galaxy on  heating and cooling
of the ICM.
%\cite{monograph}.

\section{X-ray observations}
\label{secbel:2}
Cygnus A was observed with \xmm\ in October 2005 in two separate exposures of 22 ks  (ObsIDs 0302800101) and 19 ks (ObsIDs 0302800201) respectively. %.
In this preliminary work we use local background for the analysis of the central 5 arcminutes, while the X-ray background is modelled for the large scale structure analysis using blank-sky background.

In this paper we also present a new analysis of the \cha\ data. \cha\ observed Cyg A in different exposures ranging from 0.8 to 59 ks, for a total combined exposure of 213 ks. Only one of the observations was obtained with the ACIS-S as primary instrument (Obs ID 360). Here we only use 120 ks, by excluding the shorter exposures and the ACIS-S observation for simplicity. 

\section{Heating}

\cite{smithetal02} and \cite{wsy06} used 20 and 50 ks \cha\ observations to measure variations of temperature across the edge-brightened structure suggesting the detection of the shock of Mach number 1.3 associated with the cocoon as expected in models of radio bubble expansion (e.g. \cite{bc89}). This implies that the simple model in which the expansion of the radio lobes is able to heat the surrounding gas, with the effect of a shock at the edge of the cocoon, is reasonably verified for Cygnus A. The analysis of \cite{wsy06} measures a temperature of 6 keV within the cocoon and a temperature of 4.6($\pm0.5$) keV outside it (see Figure 3 of \cite{wsy06}).

\subsection{Results}
We used the higher sensitivity of \xmm\ to try to verify the \cha\
results.
In both \cha\ and \xmm\ observations there is a rather sharp surface brightness edge at
the location of the possible cocoon, which is more clear in the \cha\ image thanks to the 1 arcsec Point Spread Function (PSF).

\begin{figure}
\centering
\includegraphics[height=5.cm]{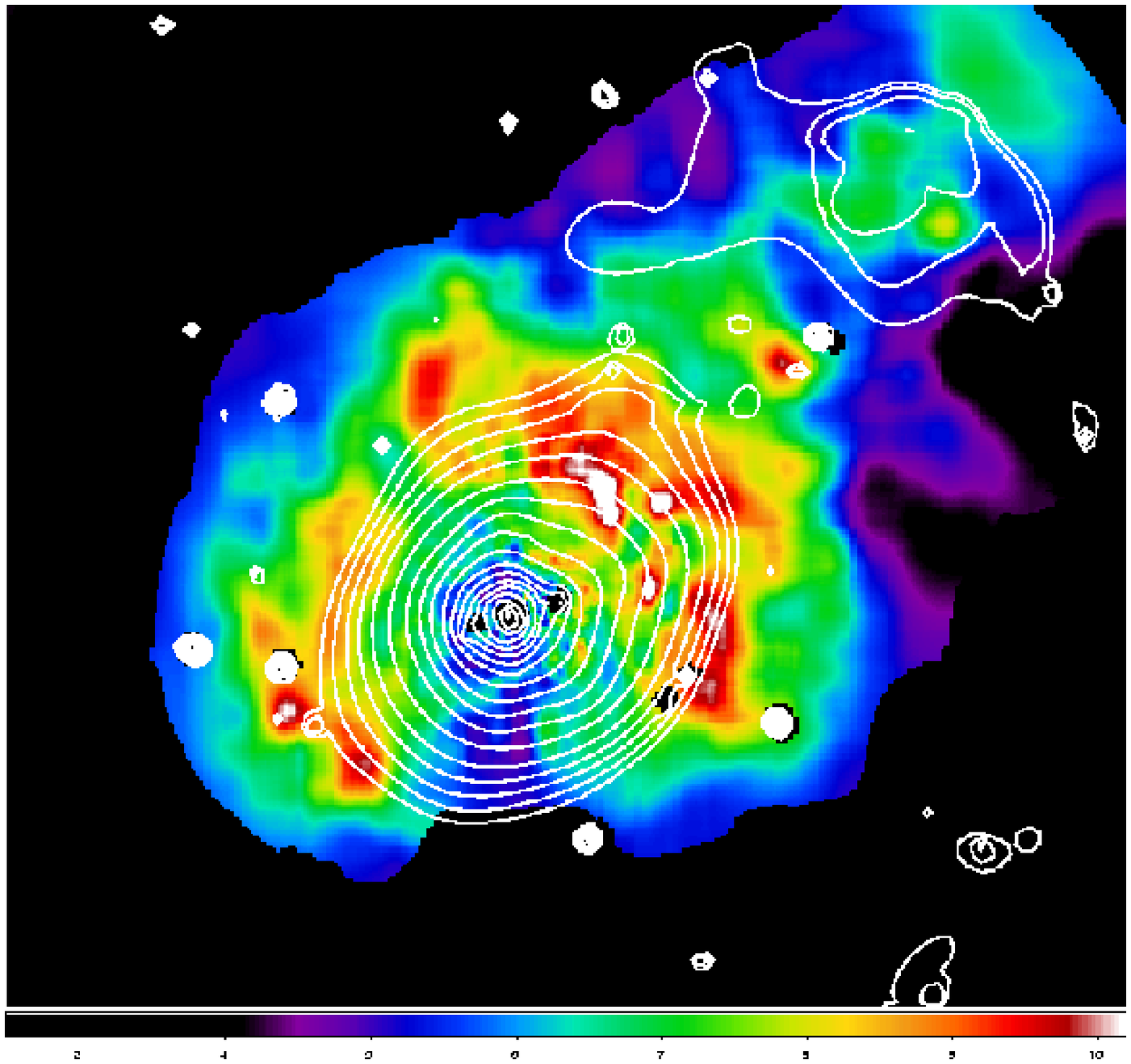}
\includegraphics[height=4.cm]{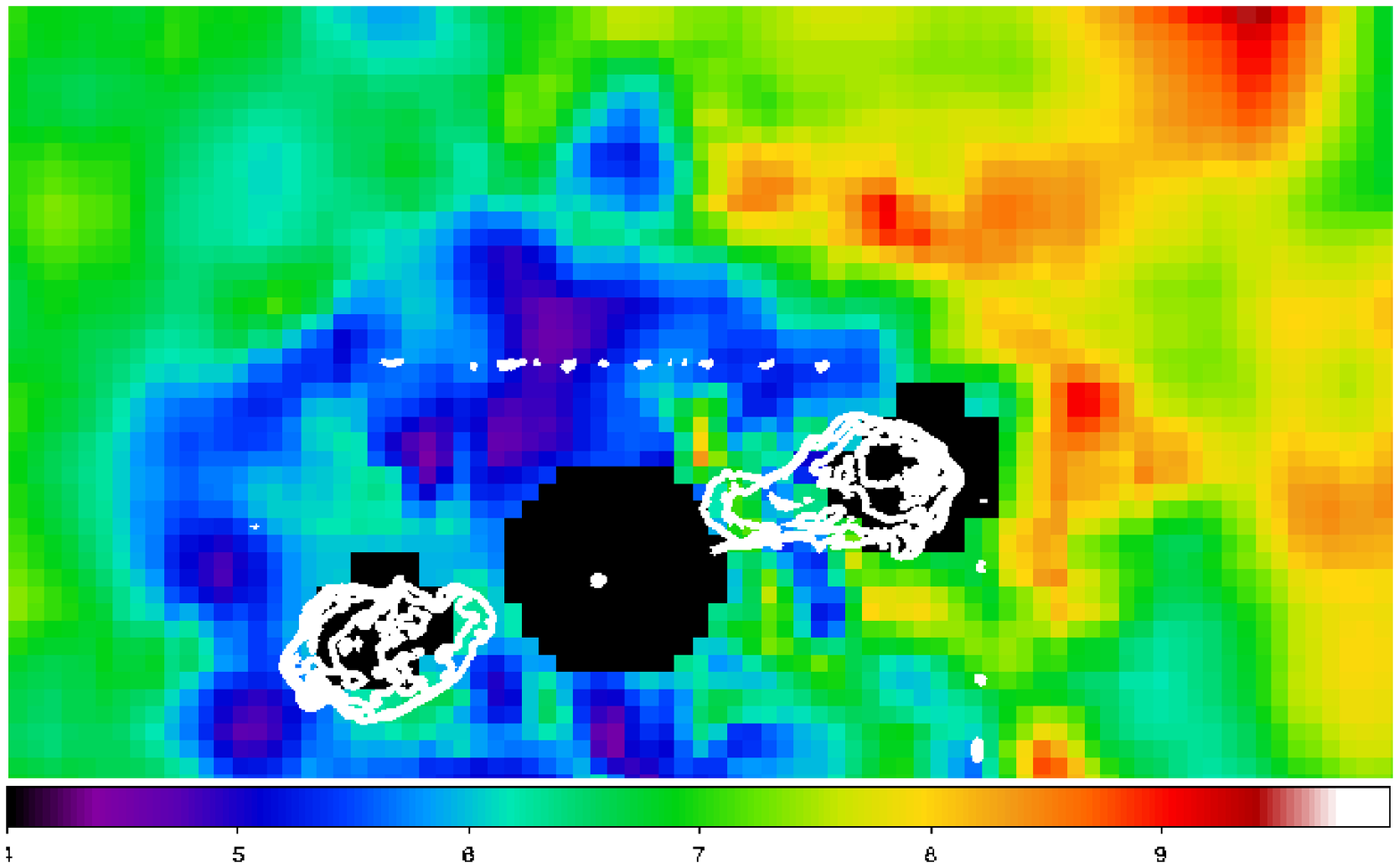}
%\picplace{5cm}{2cm} % Give the correct figure height and width in cm
%
\caption{Left: Temperature map of the Cygnus cluster generated with the wavelet algorithm described in \cite{bourdin04}. Contours are the X-ray surface brightness in the energy range 0.5-7.0 keV. Right: zoom on the temperature map in the central 5 arcmin. Contours are from the 1.4 GHz radio map.}
\label{fig:tmaplarge}       % Give a unique label
\end{figure}

We generated a temperature map of Cygus A using the wavelength algorithm
described in \cite{bourdin04}. (The new version of this method 
includes a model of the X-ray background in addition to the particle background). Figure
\ref{fig:tmaplarge} shows the temperature map of the Cygnus A cluster
in the whole field of view of \xmm.
The large scale structure is clearly complex and shows regions of high
temperature gas, mostly associated with the merging of the main
cluster with the object detected to the NW. We will discuss
the merging event in detail in Belsole et al. (in prep.).
In Figure \ref{fig:tmaplarge}, right, we show the temperature distribution of
the central 5 arcmin. The contours are the radio emission at 1.4
GHz. The core and hotspots were masked while generating the
temperature map and thus appear as holes in the figure.

\begin{figure}
\centering
\includegraphics[height=4cm]{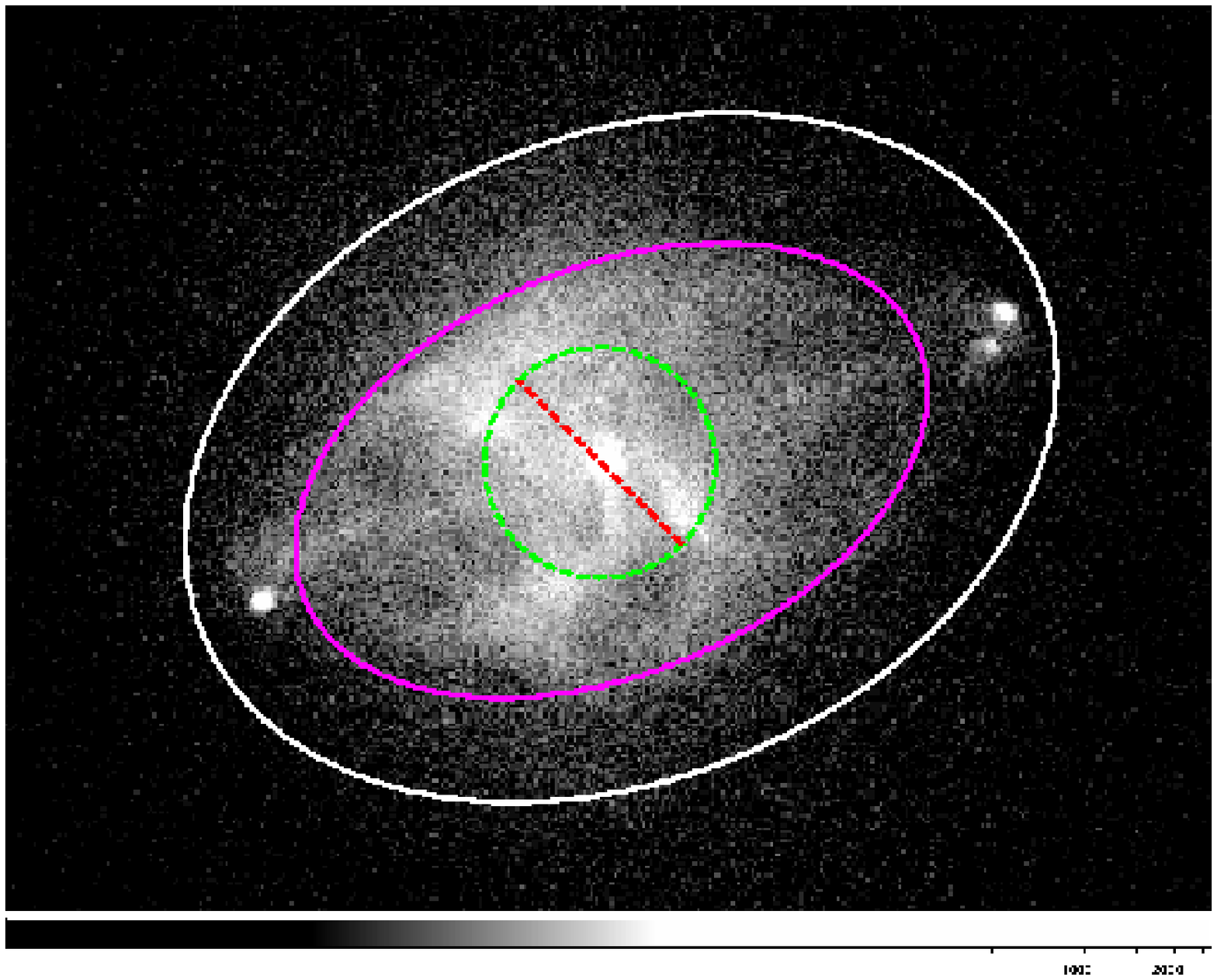}
\includegraphics[height=4cm]{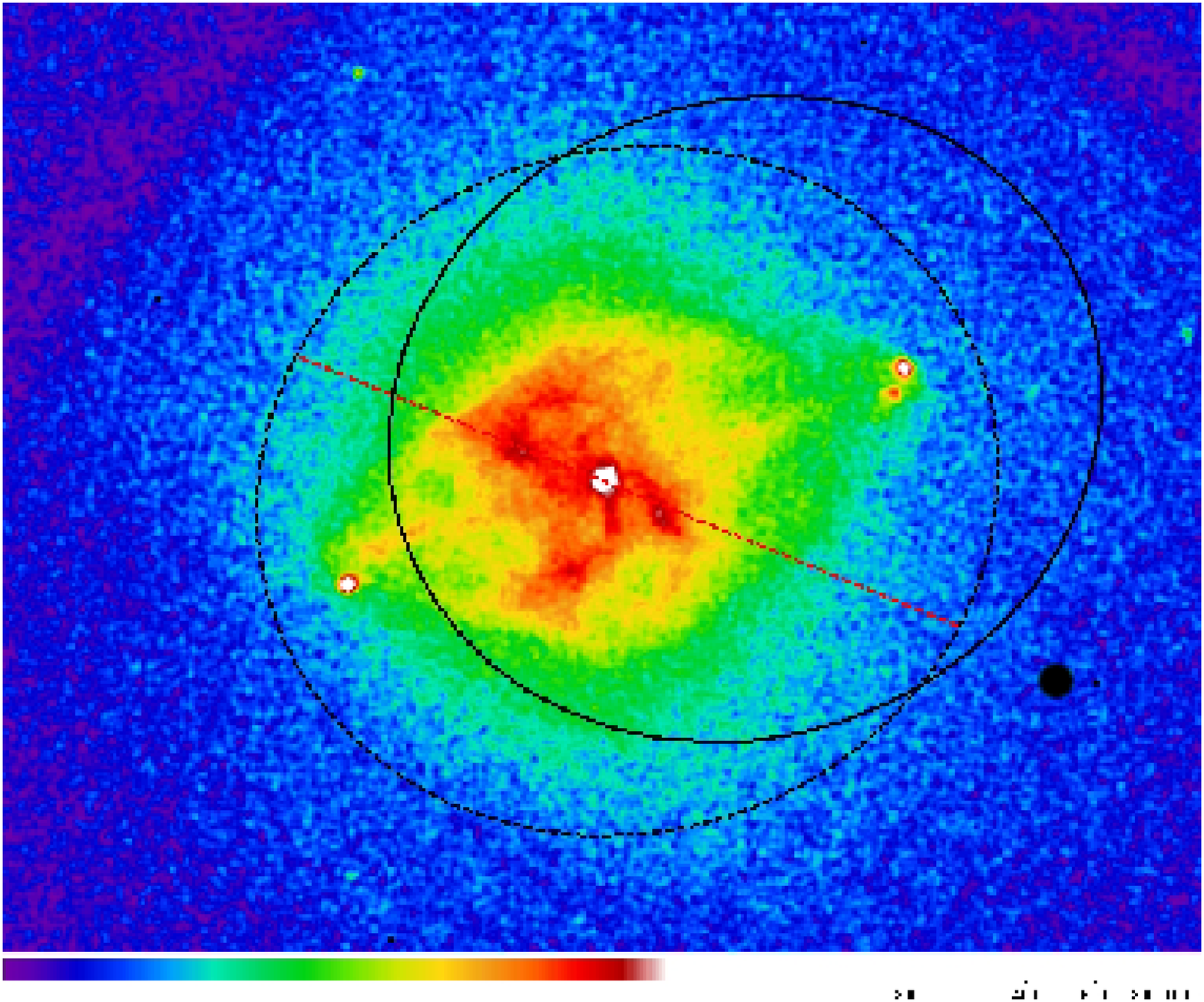}
%\picplace{5cm}{2cm} % Give the correct figure height and width in cm
%
\caption{Left: Chandra image in the 0.5-7.0 keV energy band. The elliptical regions indicate the areas used to extract the spectrum inside and outside the cocoon. Right: zoom of the region used to evaluate the temperature of the bow shock detected at the edge of the NW hotspot. This should be compared with Figure 1, right.}
\label{fig:cocoon1}       % Give a unique label
\end{figure}

Figure \ref{fig:cocoon1} left, shows a combined \cha\ image in the energy
band 0.5-7.0 keV. The superimposed elliptical regions were used to evaluate the
temperature inside and outside the cocoon with \xmm.
We measure a
temperature inside  the cocoon (inner ellipse) of $4.68\pm0.12$ keV and
a temperature in the elliptical annulus just outside the cocoon of
$4.51\pm0.2$ keV. The temperature of the ICM further out is found to be around 7 keV.

We observe that there is an arc-shaped region at the termination of the
NW lobe (Fig. \ref{fig:tmaplarge}, right). This region is found to be at k$T>12$ keV,  
which is significantly higher than the temperature of  
the  surrounding areas both closer to  the centre (at the location of
the hotspot and within the cocoon) and further out.

\subsection{Discussion}
The higher sensitivity of \xmm\ allows a more precise measurement of
the temperature, but some features may be washed out because of the PSF. 
The surface brightness edge observed neatly with \cha\ is not as sharp with \xmm\ and the radial profile does not show any obvious discontinuity.
At a distance of $\sim 2.5$ arcmin from the centre, we detect a hot, k$T$=[9-11] keV,  horse-shoe-shape
 area that we associate with the merging event. This may partially
explain why we measure a higher temperature far from the cocoon.
However, the temperature of the {\it belts} (Fig. \ref{fig:cocoon1}) is significantly cooler
than the immediate surrounding, in agreement with \cite{wsy06}.  With the \xmm\ data we
show that the temperature inside and outside the cocoon are in fact in
agreement. This is not in disagreement with the results of  \cite{wsy06}, as once statistical errors in their analysis are
taken into account, the two temperatures they measure  (inside and outside the
cocoon) are in fact consistent. 

The arc-shaped hot region may be associated with the bow shock due to the expansion of the radio lobes in the NW direction. Assuming Rankine-Hugoniot jump conditions, the shock (12.8$\pm2.3$ keV) and pre-shock (8.6$\pm0.6$ keV) temperature yield a Mach number of 1.62, or gas moving at a velocity of 2450 km s$^{-1}$. However, we are not able to explain why we observe a shock in the NW direction while the SE lobe shows a cooled region in front of it.  

\section{Cooling}

In this section we discuss a possible detection of cool gas in the centre of the Cygnus A cluster. 

The predicted mass of cool gas in ``cooling core'' clusters is inconsistent with the observed mass contained in stars and molecular gas (see e.g., McNamara and Crawford, these proceedings). The hypothesis which is mostly supported today is that the gas is prevented from cooling by some heating mechanisms that appear to be related to the AGN activity (e.g., \cite{Birzan04}, \cite{pf05} for a review). However, there is the possibility that cold material is present, but is not easy to detect or may be detected under special conditions only. Clouds of cold material can be observed by their fluorescent line at 6.4 keV, but their detection may depend on the optical depths, the shape of the emitting region and the covering fraction (e.g., \cite{White1991, Churazov98}).

\subsection{Results}
\begin{figure}
\centering
\includegraphics[height=4.9cm,angle=-90]{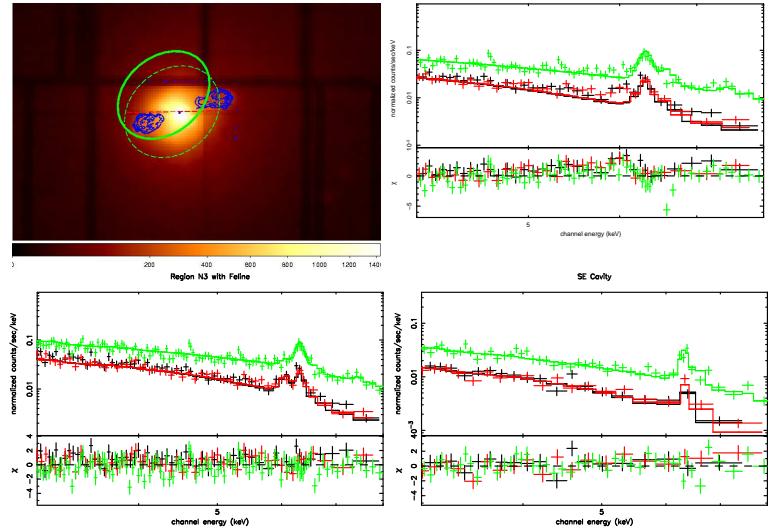}
\includegraphics[height=5cm,angle=-90]{fig3rt.ps}
\includegraphics[height=5cm,angle=-90]{fig3lb.ps}
\includegraphics[height=5cm,angle=-90]{fig3rb.ps}
%\picplace{5cm}{2cm} % Give the correct figure height and width in cm
%
\caption{Top left: The region (N3) superimposed on the epic-pn image is one of the crescent-shape area used for the spectral analysis described above. It is at an average distance of 65 arcsec from the centre; Top right: spectrum of region N3 fitted with a mekal model. The residuals at the position of the Fe neutral line is evident. Bottom left: spectrum of region N3 fitted with a mekal+power law+neutral 6.4 keV line. Bottom right: spectrum of a region at a mean distance of 40 arcsec corresponding to the SE lobe cavity. The spectrum is well fitted with a simple mekal model of k$T\sim 5$ keV. }
\label{fig:cooling}       % Give a unique label
\end{figure}

We extracted spectra of the central region of Cygnus A cluster at increasing distance from the centre and in different directions: taking as  the axis of symmetry the jet line, spectra were extracted in 3 elliptical annular sectors (crescent-shape regions) in the direction orthogonal to the jet and at a mean distance of 25, 50, and 65 arcsec, to the NE and the SW. Two more spectra were extracted in the cavities occupied by the radio lobes. Figure \ref{fig:cooling} shows the most distant annular sector to the NE. The spectrum of this region is fitted with a thermal model of temperature k$T= (6.52\pm0.16$) keV. The fit is not excellent and the reduced $\chi^2$ of 1.4 is due to a significant residual at the position of the neutral Fe line at 6.4 keV (Figure \ref{fig:cooling}, top left). The addition of a power law and a neutral 6.4 keV line (of equivalent width 210 eV)  to the model significantly improves the fit and accounts for the residual at the Fe line position (Figure \ref{fig:cooling}, bottom right).

The spectrum obtained in the SE lobe cavity (Figure \ref{fig:cooling}), which is at a slightly closer distance of 40 arcsec to the nucleus, does not show the emission line at 6.4 and is adequately fitted with a thermal model at k$T \sim 5$ keV (we also account for a possible inverse-Compton component associated with the radio lobes).

\subsection{Discussion}
The results described above suggest the presence of cold gas observed through its  fluorescent line at 6.4 keV \cite{Churazov98}. The \xmm\ PSF is a major concern here as the nucleus spectrum also has a strong neutral Fe line emission. We will discuss in detail the core spectrum in a separate paper. 
However, the \xmm\ PSF is rather circular (with the MOS 2 camera showing a more triangular shape), and it would be difficult to explain why we detect Fe 6.4 keV  line emission from a region at 65 arcsec to the NE of the nucleus but not from a closer region in the jet direction, corresponding to the cavity of the SE radio lobe. Inspection of  \cha\ data is a  natural choice to confirm our detection. Using the ACIS-I  observations, we generated a merged image in the energy bands 5335-5665 eV (A), 5820-6150 (B, or the Fr 6.4 keV line), 6180-6410 eV (C, or the Fe 6.7 keV line), and 6835-7165 eV (D). We then interpolated the continuum under the lines using images A and D, and  by subtracting the continuum to the line image, we generated the images of the neutral (fluorescent) Fe line and the ionised (ICM) Fe K line in order to localise the regions with significant  Fe line emission. We find that there are several, patchy regions where the neutral line emission is detected at $\sim 4$ sigma above the background and some of these regions correspond to a lack of emission at 6.7 keV. However, a precise spectral analysis is not possible with the low photon statistics (we will discuss these results in a forthcoming paper). The areas which shows the patchy 6.4 keV emission are at a distance from the nucleus which is still affected by the \xmm\ PSF, so no conclusive answer can yet be drawn from this analysis. However, we find that the luminosity of the neutral emission line in region N3 is consistent with the expected ratio between the fluorecent line and the K-edge \cite{kk87} for reflecting clouds illuminated by the AGN (assuming a covering fraction of $\sim 3\%$ at their distance from the core). We notice that the global spectrum described in \cite{smithetal02}, which excludes the nuclear area, also shows excess emission at the position of the 6.4 emission line (see Figure 7 of \cite{smithetal02}). However, the authors do not discuss their residuals. 
A deeper investigation of this detection is needed using both \xmm\ and \cha. 

\section{Summary and preliminary conclusions}
We have presented new \xmm\ data of Cygnus A and we have re-analysed 120 ks \cha\ observations. We find that the ICM temperature distribution of the cluster hosting the radio galaxy Cygnus A is complex and at least some of this complexity is due to the merging event with a secondary cluster detected to the NW.
With both \xmm\ and \cha\ we detect an elliptically-shaped sharp edge in the gas surface brightness, with a major axis similar to the distance between the two hotspots. This is interpreted as the predicted cocoon of shocked gas due to the expansion of the radio lobes. However no significantly different temperature is detected inside and outside the cocoon. 

We may have detected a shock associated with the expansion of the lobes along the major axis at the location of the NW lobe, with a bow shock of Mach number 1.6, which may be a lower limit due to the projected hot gas at large scale. However we are unable to explain the apparent cold gas associated with the edge of the SE lobe.

Using \xmm\, we may have spectrally detected excess emission at 6.4 keV in the preferred axis orthogonal to the jet direction. This may be due to fluorescent line from geometrically small, optically thin cold clouds reflecting photons from the core and the hot gas in the cocoon. Although contamination of the Cygnus A core due to the \xmm\  PSF is likely, this cannot explain the axis-asymmetry we observe, since in the direction of the jet no 6.4 keV emission line is detected.

\section{Acknowledgements}
We are very grateful to the local committee for this very well organised conference with plenty of interesting results. We also thank Giovanni Miniutti and Gabriel W. Pratt for useful discussion, Herv\'e Bourdin for providing a user-friendly version of the temperature map algorithm and Roderick Johnstone for technical support.

%\begin{equation}
%\vec{a}\times\vec{b}=\vec{c}
%\end{equation}

%\subsubsection{Subsubsection Heading}
%Your text goes here. Use the \LaTeX\ automatism for cross-references as
%well as for your citations, see Sect.~\ref{sec:1}.

%\paragraph{Paragraph Heading} %
%Your text goes here.

%\subparagraph{Subparagraph Heading.} Your text goes here.%
%
%\index{paragraph}
% Use the \index{} command to code your index words
%\begin{table}
%\centering
%\begin{tabular}{lll}
%\hline\noalign{\smallskip}
%\end{tabular}
%\end{table}
%\begin{figure}
%\centering
%\includegraphics[height=4cm]{figure.eps}
%\picplace{5cm}{2cm} % Give the correct figure height and width in cm
%\end{figure}
%
% For built-in environments use

% BibTeX users please use
% \bibliographystyle{}
% \bibliography{}

\begin{thebibliography}{99.}
\bibitem {arnaud84} Arnaud, K.A., Fabian, A.C., Eales, S.A., Jones, C., Forman, W., 1984, MNRAS, 211, 981
\bibitem {bc89} Begelman, Mitchell C., Cioffi, Denis F.,  1989, ApJ, 345, 21
\bibitem {Birzan04} Bir\^zan L., Rafferty, D.A., McNamara, B.R., Wise, M.W., Nulsen, P.E.J., 2004, ApJ, 607, 800
\bibitem {bourdin04} Bourdin, H., Sauvageot, J.-L., Slezak, E., Bijaoui, A.; Teyssier, R., 2004, A\&A, 414, 429
\bibitem {carillietal94}Carilli, C. L., Perley, R. A., Harris, D. E., 1994, MNRAS, 270. 173
\bibitem {Churazov98} Churazov, E., Sunyaev, R., Gilfanov, M., Forman, W., Jones, C., 1998, MNRAS, 297, 127
\bibitem {hcp94} Harris, D.E., Carilli, C.L., Perely, R.A., 1994, Nat., 367, 713
\bibitem {kk87} Krolik, J.H., Kallman, T. R., 1987, 320, L5
\bibitem {lom05}Ledlow, M.J., Owen, N.O., Millar, N.A., 2005, AJ, 130, 47
\bibitem {msv99}Markevitch, M., Sarazin, C.L., Vikhlinin, A., 1999, ApJ, 521, 526
\bibitem {owenetal97}Owen, F.N., Ledlow, M.J., Morrison, G.E., Hill, J.M., 1997, ApJ, 488, L15
\bibitem {pf05} Peterson, J.R., Fabian, A.C., 2005, 
\bibitem {rf96} Reynolds, C.S, Fabian, A.C., 1996, MNRAS, 278, 479
\bibitem {smithetal02}Smith, D.A., Wilson, A.S., Arnaud, K.A., Terashima, Y., Young, A.J., 2002,  ApJ, 565, 195
\bibitem {wys00} Wilson, A.S., Young, A.J., Shopbell, P.L., 2000, ApJ, 544, L27
\bibitem {wsy06} Wilson, A.S.,Smith, D.A., Young, A.J., 2006, ApJ, 644, L9
\bibitem {youngetal02}Young, A.J., Wilson, A.S., Terashima, Y., Arnaud, K.A. Smith, D.A., 2002, ApJ, 564, 176
\bibitem {White1991} White, D. A., Fabian, A. C., Johnstone, R. M., Mushotzky, R. F., Arnaud, K. A., 1991, MNRAS, 252, 72
\end{thebibliography}
%
% Non-BibTeX users please follow the syntax
% the syntax of "referenc.tex" for your own citations
%\input{referenc}
%%%%%%%%%%%%%%%%%%%%%%%%%%%%%%%%%%%%%%%%%%%%%%%%%%%%%%%%%%%%%%%%%%%%%%  }

%%%%%%%%%%%%%%%%%%%%%%%%%%%%%%%%%%%%%%%%%%%%%%%%%%%%%%%%%%%%%%%%%%%%%%

%\printindex
\end{document}